\documentclass[useAMS,referee, usenatbib]{biom}
\usepackage[plain,noend]{algorithm2e}
\def\bSig\mathbf{\Sigma}

\newcommand{\ind}{\perp\!\!\!\!\perp} 
\usepackage{amsmath}
\DeclareMathOperator*{\argmax}{arg\,max}
\DeclareMathOperator*{\argmin}{arg\,min}
\newcommand{\norm}[1]{\lVert#1\rVert}
\newcommand\muko{\mu_{k}^{(0)}}

\usepackage{times}
\usepackage{graphicx}
\usepackage{bm}
\usepackage{changepage}
\usepackage{graphicx}
\usepackage{enumerate}
\usepackage{amssymb}
\usepackage{subcaption}
\usepackage{url} 
\usepackage{comment}

\usepackage[colorlinks=true,linkcolor=blue,citecolor=blue]{hyperref}

\title[Multi-Study $R$-Learner]{Multi-Study $R$-Learner for Estimating Heterogeneous Treatment Effects Across Studies Using Statistical Machine Learning}

\author
{Cathy Shyr\emailx{cathy.shyr@vumc.org} \\
Department of Biomedical Informatics, Vanderbilt University Medical Center, \\
2525 West End Avenue, Nashville, Tennessee 37203, U.S.A.
\and
Boyu Ren\emailx{bor158@mail.harvard.edu} \\
Laboratory for Psychiatric Biostatistics, McLean Hospital, \\
115 Mills St., Belmont, Massachusetts 02478, U.S.A.
\and
Prasad Patil \emailx{patil@bu.edu}\\
Department of Biostatistics, Boston University School of Public Health, \\
715 Albany St., Boston, Massachusetts 02115, U.S.A.
\and
Giovanni Parmigiani \emailx{gp@jimmy.harvard.edu}\\
Department of Data Science, Dana-Farber Cancer Institute,\\
450 Brookline Ave., Boston, Massachusetts 02215, U.S.A.
}

\begin{document}


\date{}



\pagerange{\pageref{firstpage}--\pageref{lastpage}} 
\volume{}
\pubyear{}
\artmonth{}


\doi{}


\label{firstpage}

\begin{abstract}
Estimating heterogeneous treatment effects (HTEs) is crucial for precision medicine. While multiple studies can improve the generalizability of results, leveraging them for estimation is statistically challenging. Existing approaches often assume identical HTEs across studies, but this may be violated due to various sources of between-study heterogeneity, including differences in study design, study populations, and data collection protocols, among others. To this end, we propose a framework for multi-study HTE estimation that accounts for between-study heterogeneity in the nuisance functions and treatment effects. Our approach, the multi-study $R$-learner, extends the $R$-learner to obtain principled statistical estimation with machine learning (ML) in the multi-study setting. It involves a data-adaptive objective function that links study-specific treatment effects with nuisance functions through membership probabilities, which enable information to be borrowed across potentially heterogeneous studies. The multi-study $R$-learner framework can combine data from randomized controlled trials, observational studies, or a combination of both. It's easy to implement and flexible in its ability to incorporate ML for estimating HTEs, nuisance functions, and membership probabilities.  In the series estimation framework, we show that the multi-study $R$-learner is asymptotically normal and more efficient than the $R$-learner when there is between-study heterogeneity in the propensity score model under homoscedasticity. We illustrate using cancer data that the proposed method performs favorably compared to existing approaches in the presence of between-study heterogeneity.  
\end{abstract}

%



\maketitle
%

\section{Introduction}
\label{s:intro}

Heterogeneous treatment effect (HTE) estimation is central to modern statistical applications such as precision medicine \citep{collins2015new}. In practice, patients with different characteristics can respond differently to a given treatment. The overall treatment effect, which is typically reported in randomized controlled trials (RCTs), may not be directly applicable to all patients in clinical practice \citep{dahabreh2023toward}. HTEs vary across individuals based on their unique characteristics and, as a result, are crucial in informing personalized treatment strategies \citep{kosorok2019annual}. Recently, facilitation of systematic data sharing led to increased access to multiple studies that target the same treatment, providing opportunities for improving the accuracy, precision, and generalizability of treatment effect estimation \citep{ manzoni2018genome}. Despite this, estimating treatment effects on multiple studies is statistically challenging due to various sources of between-study heterogeneity, including differences in study design, data collection methods, sample characteristics, among others. 

Recently, there is growing interest in combining data from multiple studies to estimate treatment effects (\cite{degtiar2023review, colnet2024causal}). A common assumption in this literature is ignorability of study label given covariates \citep{hotz2005predicting, stuart2011use, tipton2013improving, hartman2015sample, buchanan2018generalizing, kallus2018removing, egami2021covariate, colnet2024causal}. Mathematically, this assumption states that the potential outcome $Y(a)$ under treatment $a \in \mathcal{A}$ is independent of the study label $S \in \{1, \ldots, K\}$ given the covariates $X \in \mathcal{X}$, i.e., $Y(a) \ind S \mid X$. An important implication of this assumption is transportability or mean exchangeability of the HTEs \citep{dahabreh2019efficient, dahabreh2019extending, wu2021integrative, colnet2024causal}. That is, $E[Y(a) \mid X =x, S = k] = E[Y(a) \mid X=x, S = k'] = E[Y(a) \mid X=x]$ for all studies $k \neq k'$, treatment $a \in \mathcal{A}$, and covariates $x \in \mathcal{X}$. In practice, however, this may be untenable due to various sources of between-study heterogeneity. Transportability of the HTEs will be violated if a treatment effect modifier was not measured across all studies due to differences in study design or data collection protocols. Another example is when the HTEs differ due to the heterogeneity in study populations (e.g., differences in the distribution of treatment effect modifiers across studies).

We propose a statistical machine learning (ML) framework for estimating HTEs on multiple studies without assuming transportability. Adapting statistical ML methods for HTE estimation is attractive because of their flexibility and strong empirical performance \citep{brantner2023methods}. Notable examples in the single-study setting include tree-based approaches \citep{wager2018estimation}, boosting \citep{powers2018some}, neural networks \citep{shalit2017estimating}, and Lasso \citep{imai2013estimating}. Despite growing interest in these adaptations, developing them can be labor-intensive. Moreover, they generally do not have theoretical guarantees for improvement in isolating causal effects compared to simple nonparametric regressions. To this end, \cite{nie2021quasi} proposed the $R$-learner, a framework that is not only algorithmically flexible, allowing any off-the-shelf ML method to be employed, but also quasi-oracle in the case of penalized kernel regression. We extend the $R$-learner to account for between-study heterogeneity in the multi-study setting.

Our paper makes several contributions. 1) The proposed framework, the multi-study $R$-learner, is robust to between-study heterogeneity in the nuisance functions and HTEs. It involves a data-adaptive objective function that links study-specific treatment effects with nuisance functions through membership probabilities. These probabilities enable cross-study learning, thereby allowing information to be borrowed across heterogeneous studies. \cite{nie2021quasi}'s $R$-learner is a special case of the proposed approach in the absence of between-study heterogeneity. 2) Under homoscedasticity, we show analytically that the multi-study $R$-learner is asymptotically unbiased and normally distributed in the series estimation framework. In the two-study setting, the proposed method is more efficient than the $R$-learner when there is between-study heterogeneity in the propensity score models. 3) Results from extensive evaluations using cancer data showed that the multi-study $R$-learner performs favorably compared to other methods as between-study heterogeneity increases. 4) The multi-study $R$-learner is easy to implement and allows flexible estimation of nuisance functions, HTEs, and membership probabilities using modern ML techniques. 5) It can be used to combine RCTs, observational studies, or a combination of both.

 In this paper, there are two concepts of heterogeneity: heterogeneity in the treatment effect (i.e., HTE) across individuals and heterogeneity of various distributions between studies. The former refers to variation in the treatment effect across individuals with different covariates. For example, the effect of a targeted cancer treatment can vary across different cancer sub-types depending on whether the genetic alteration targeted by the drug is present in the sub-type \citep{habeeb2016use}. In contrast, between-study heterogeneity refers to differences in study-level characteristics. These concepts are distinct in that HTEs can occur within studies in the absence of between-study heterogeneity. They are also connected, as the presence of between-study heterogeneity may result in HTEs that differ across studies. 

The rest of the paper is organized as follows: Section \ref{s:methods} formalizes the problem setup and introduces the multi-study $R$-learner framework, Sections \ref{s:theory} and \ref{s:simulations} provide the theoretical and simulation results, respectively, and Section \ref{s:data_app} illustrates our approach in a breast cancer data application. 

\section{Methods}
\label{s:methods}
\subsection{Problem Setup}
Suppose we have data from $K$ studies, indexed by $k = 1,\ldots, K$. For individual $i = 1, \ldots, n$, we observe the random tuples $(Y_i, X_i, A_i, S_i)$ where $Y_i \in \mathbb{R}$ denotes the outcome, $X_i \in \mathcal{X} \subset \mathbb{R}^p$ the baseline covariates, $A_i \in \{0, 1\}$ the treatment assignment, and $S_i\in \{1,\ldots,K\}$ the study label drawn randomly according to $p(k\mid x) = \text{pr}(S_i = k \mid X_i = x)$ for study $k = 1, \ldots, K$. We refer to $p(k\mid x)$ as the membership probability of belonging to study $k$ given covariate $x$. Study $k$ consists of $n_k$ independent and identically distributed tuples, and $n = \sum_{k=1}^K n_k$ is the combined sample size. We adopt the potential outcomes framework \citep{rubin1974estimating} and let $\{Y_i(1), Y_i(0)\}$ denote the potential outcomes that would have been observed given the treatment assignments $A_i = 1$ and $A_i = 0,$ respectively. We now define study-specific HTEs for the multi-study case. 
\begin{definition}\label{def1} The study-specific HTE, given covariates $x \in \mathcal{X}$, in study $k$ is 
$$\tau_k(x) = E[Y_i(1) - Y_i(0)\mid X_i = x, S_i = k], \;\;\;\;\;\; i = 1, \ldots, n \;\;\;\;\;\; k = 1, \ldots, K.$$
\end{definition}

\noindent Our goal is to estimate the overall HTE by leveraging data from the $K$ studies. We refer to this simply as HTE without the study-specific designation and denote it as $\tau(x)$.

\begin{definition}\label{def2} The HTE given covariates $x \in \mathcal{X}$ is 
    $$\tau(x) = E[Y_i(1) - Y_i(0)\mid X_i = x], \;\;\;\;\;\; i = 1, \ldots, n.$$
\end{definition}

\noindent From Definitions~\ref{def1} and~\ref{def2}, it follows that 
$\tau(x) = \sum_{k=1}^K \tau_k(x)p(k\mid x).$ In practice, $\tau(x)$ has important implications for informing personalized treatment strategies, as clinicians may want to understand the effect of a treatment for patients with covariates $x$ based on information from $K$ related but somewhat heterogeneous studies. Because our causal estimand of interest $\tau(\cdot)$ is a weighted average of $\tau_k(\cdot)$, $k = 1, \ldots, K$, we proceed by estimating $\tau_k(\cdot)$ from the $K$ studies and consider properties of our procedure for this estimation step as well. To this end, we make the following identifiability assumptions:

\begin{assumption}[Consistency]\label{assumption1}
    $Y_i = Y_i(1)A_i + Y_i(0)(1 - A_i)$ \hspace{0.5em} for $i = 1, \ldots, n.$
\end{assumption}

\begin{assumption}[Mean unconfoundedness within study]\label{assumption2}
    $E[Y_i(a)\mid A_i=a, X_i=x, S_i = k] = E[Y_i(a)\mid X_i=x, S_i = k],$ for $ k=1, \ldots, K $, $a \in \{0, 1\}$, and $i = 1, \ldots, n$.
\end{assumption}

\begin{assumption}[Positivity of treatment within study]\label{assumption3}
    $0 < \text{pr}(A_i = a \mid X_i = x, S_i = k) < 1$ for $ k=1, \ldots, K $, $a \in \{0, 1\}$, and $i = 1, \ldots, n$.
\end{assumption}

\noindent Assumption~\ref{assumption1} states that the observed outcome is equal to the potential outcome under the treatment actually received. 
Assumption \ref{assumption2} posits that there is no unmeasured confounding in the mean outcome function conditional on the observed covariates within each study. This is less restrictive than  $\{Y(1), Y(0)\} \ind A \mid X, S$ and still leads to the same optimization problem based on the multi-study $R$-loss, which we formally define in Section \ref{sec:msr-learner}. Assumption \ref{assumption3} states that it's possible for individuals to receive either treatment in any study.

\subsection{Multi-study Robinson's Transformation}
To motivate the multi-study $R$-learner, we extend Robinson's transformation \citep{robinson1988root} to the multi-study setting. To begin, let $m(x) = E[Y_i \mid X_i=x]$ denote the conditional mean outcome function. It's helpful to first re-write $m(x)$ as $\sum_{k=1}^K \{\muko(x) + e_k(x)\tau_k(x)\}p(k \mid x),$
where $\muko(x) = E[Y_i(0)\mid X_i = x, S_i = k]$ denotes the study-specific conditional response under treatment $a = 0$, and $e_k(x) = \text{pr}(A_i = 1 \mid X_i = x, S_i = k)$ the study-specific propensity score function. 
We let
$
\epsilon_i = Y_i - \sum_{k=1}^K \left\{\muko(X_i) + A_i\tau_k(X_i)\right\}p(k \mid X_i)
$
denote the error term. Then, it follows that
\begin{equation}
    \label{msRlearner}
    Y_i - m(X_i) = \sum_{k=1}^K \{A_i - e_k(X_i)\}\tau_k(X_i)p(k \mid X_i) + \epsilon_i, \hspace{1em} i = 1, \ldots, n.
\end{equation}
The function $m(X_i)$ can be re-written as $\sum_{k=1}^K p(k\mid X_i)m_k(X_i)$, where $m_k(X_i) = E[Y_i \mid X_i = x, S_i = k]$ is the conditional mean outcome function in study $k$. We establish the validity of Eqn (\ref{msRlearner}) as the multi-study Robinson's transformation in Proposition \ref{proposition1}.
\begin{proposition}
\label{proposition1}
Under Assumption~\ref{assumption2}, $E[\epsilon_i\mid A_i, X_i] = 0.$
\end{proposition}

\noindent A proof is provided in the Supplementary Material. In the absence of between-study heterogeneity in the propensity score functions or HTEs, Eqn (\ref{msRlearner}) is identical to the single-study Robinson's transformation, 
$$
    Y_i - m(X_i) = (A_i - e(X_i))\tau(X_i) + \epsilon_i,
$$
with error $\epsilon_i = Y_i-\{\mu^{(0)}(X_i)+A_i\tau(X_i)\},$ where $\mu^{(0)}(X_i) = E[Y(0)\mid X_i]$. Thus, \cite{nie2021quasi}'s $R$-learner is a special case of the multi-study $R$-learner, which we formally introduce in the next section.

\subsection{Multi-study $R$-Learner}
\label{sec:msr-learner}
The multi-study Robinson's transformation in Eqn (\ref{msRlearner}) implies a duality between the estimation of the study-specific HTEs $\tau_1(\cdot), \ldots, \tau_K(\cdot)$ and the regression problem with outcome $Y - m(X)$ and covariate $\{A - e_k(X)\}p(k\mid X)$ for $k=1, \ldots, K$. Consider the multi-study oracle $R$-loss, which is defined for fixed nuisance functions and membership probabilities as
\begin{equation}
    \label{oracle_loss}
      L_n\left(\tau_1(\cdot), \ldots, \tau_K(\cdot)\right) = \frac{1}{n}\sum_{i=1}^n\left[\{Y_{i}-m(X_{i})\}-\sum_{k=1}^K\{A_{i}-e_k(X_{i})\}p(k \mid X_i)\tau_k(X_{i})\right]^2.
\end{equation}
In practice, the true nuisance functions and membership probabilities, i.e., $m(\cdot)$, $e_k(\cdot)$, and $p(k\mid \cdot)$, are unknown and need to be estimated from data to obtain $\hat m(\cdot), \hat e_k(\cdot)$, and $\hat p(k \mid \cdot)$. This can be done using any ML technique appropriate for the problem at hand, such as penalized regression, deep neural networks, or boosting. 
Next, to mitigate potential overfitting and ensure identifiability, we incorporate a regularizer $\Lambda_{\tau_k}$ on the complexity of $\tau_k(\cdot)$. 
Putting these pieces together, the optimal study-specific HTEs 
satisfy: $\{\hat{\tau}_1(\cdot), \ldots, \hat{\tau}_K(\cdot)\} = $ 
 \begin{equation}
     \label{multiRloss}
       \argmin_{{\tau}_1, \ldots, {\tau}_K} \left\{\frac{1}{n}\sum_{i=1}^n\left[\{Y_{i}-\hat{m}(X_{i})\}-\sum_{k=1}^K\{A_{i}-\hat{e}_k(X_{i})\}\hat{p}(k \mid X_i){\tau}_k(X_{i})\right]^2+\Lambda_{\tau_k}\right\}.
\end{equation}

\noindent We make the important observation that study label $S$ is implicitly incorporated in the squared error term through the estimated membership probabilities $\hat{p}(k \mid X_i)$ in Eqn (\ref{multiRloss}). Specifically, the study labels are drawn randomly according to $p(k \mid x) = \text{pr}(S_i = k \mid X_i = x)$, which encodes how likely individual $i$ with observed covariates $x$ belong in study $k$. These membership probabilities enable information to be borrowed across studies through cross-study learning. To illustrate this, consider the example where there are $K = 2$ studies measuring a single covariate (e.g., $X$ = age). We highlight two special cases. First, when there is no between-study heterogeneity in the HTEs, i.e., $\tau_1(\cdot) = \tau_2(\cdot)$, then the optimization in Eqn (\ref{multiRloss}) is equivalent to merging the studies and fitting a single $R$-learner. Second, when there is between-study heterogeneity (e.g., $\tau_1(\cdot) \neq \tau_2(\cdot)$) and no overlap between the studies' covariate distributions, i.e., individuals in study $S = 1$ have $p(1 \mid \cdot) = 1$ and those in study $S = 2$ have $p(2 \mid \cdot) = 1$, then no information is borrowed between studies. An example is when a study's exclusion criteria match the inclusion criteria of another exactly (e.g., for $X$ = age, consider a pediatric vs. adult study). These special cases represent opposite ends of the cross-study learning spectrum: homogeneous vs. entirely heterogeneous studies. Between these extremes is the case where some information is borrowed between heterogeneous studies via membership probabilities. To see this, suppose $\tau_k(X_i) = X_i\beta_k$ for $k = 1, 2$. The contributions to the squared error term in Eqn (\ref{multiRloss}) are 
$$
\begin{small}
    \begin{bmatrix}
    Y_1 - \{\hat{m}_1(X_1)\hat{p}(1\mid X_1) + \hat{m}_2(X_1)\hat{p}(2\mid X_1)\}\\
    \vdots\\
    Y_{n_1} - \{\hat{m}_1(X_{n_1})\hat{p}(1\mid X_{n_1}) + \hat{m}_2(X_{n_1})\hat{p}(2\mid X_{n_1})\}\\
    Y_{n_1+1} - \{\hat{m}_1(X_{n_1+1})\hat{p}(1\mid X_{n_1+1}) + \hat{m}_2(X_{n_1+1})\hat{p}(2\mid X_{n_1+1})\}\\
    \vdots\\
    Y_n - \{\hat{m}_1(X_n)\hat{p}(1 \mid X_n) + \hat{m}_2(X_n)\hat{p}(2\mid X_n)\}
\end{bmatrix}
\end{small}
$$
$$
\begin{small}
    - \begin{bmatrix}
    (A_1 - \hat{e}_1(X_1))\hat{p}(1\mid X_1)X_1 & (A_1 - \hat{e}_2(X_1))\hat{p}(2\mid X_1)X_1\\
    \vdots \\
    (A_{n_1} - \hat{e}_1(X_{n_1}))\hat{p}(1\mid X_{n_1})X_{n_1} & (A_{n_1} - \hat{e}_2(X_{n_1}))\hat{p}(2\mid X_{n_1})X_{n_1}\\
    (A_{n_1+1} - \hat{e}_1(X_{n_1+1}))\hat{p}(1\mid X_{n_1+1})X_{n_1+1} & (A_{n_1+1} - \hat{e}_2(X_{n_1+1}))\hat{p}(2\mid X_{n_1+1})X_{n_1+1}\\
    \vdots\\
    (A_n - \hat{e}_1(X_n))\hat{p}(1\mid X_n)X_n & (A_n - \hat{e}_2(X_n))\hat{p}(2\mid X_n)X_n
\end{bmatrix}\begin{bmatrix}
    \beta_1\\
    \beta_2
\end{bmatrix}.
\end{small}
$$
Note that each individual's contribution involves a weighted average of within-study and cross-study predictions. For example, for individuals $i = 1, \ldots, n_1$ in study 1, $\hat{m}_1(X_i)$ and $\hat{e}_1(X_i)$ represent within-study predictions, and $\hat{m}_2(X_i)$ and $\hat{e}_2(X_i)$ cross-study predictions (and vice versa for individuals in study 2). In other words, the multi-study $R$-loss incorporates cross-study learning through the estimated membership probabilities $\hat{p}(k\mid \cdot)$, linking study-specific treatment effects $\tau_k(\cdot)$ with estimated nuisance functions $\hat{m}_k(\cdot)$ and $\hat{e}_k(\cdot)$ and allowing information to be borrowed across studies. The loss function can be thought of as shrinkage towards the population average $\tau(\cdot)$, our estimand of interest, and the membership probabilities control the degree of shrinkage. In summary, the multi-study $R$-learner is a general class of algorithms defined by the following three steps: 

\begin{enumerate}
    \item[1.] Estimate study-specific nuisance functions $ e_1(\cdot), \ldots, e_K(\cdot)$ and $ m_1(\cdot), \ldots, m_K(\cdot)$ by separate analyses of the $K$ studies.
    \item[2.] Estimate membership probabilities $p(1 \mid \cdot), \ldots, p(K \mid\cdot)$ by a pooled analysis of the $K$ studies. 
    \item[3.] Estimate HTEs $\tau_1(\cdot), \ldots, \tau_K(\cdot)$ using the multi-study version of Robinson's transformation and a pooled analysis of the $K$ studies.
\end{enumerate}

To obtain estimates $\hat m(\cdot)$, $\hat e_k(\cdot)$, and $\hat p(k \mid \cdot)$, we leverage a sample-splitting strategy called cross-fitting \citep{schick1986asymptotically, chernozhukov2018double}. We randomly divide $n$ samples into $Q$ evenly-sized folds, where $Q$ is typically between 5 and 10. Let
$q(i)\in\{1,\ldots,Q\}$ denote the index of the fold that individual $i$ belongs in. We denote the estimates of $m(\cdot)$ and $p(k \mid \cdot)$ based on all samples except for those in fold $q(i)$ as $\hat{m}^{-q(i)}(\cdot)$ and $\hat{p}^{-q(i)}(k\mid \cdot)$, respectively. To obtain cross-fitted estimates of study-specific quantities such as $e_k(\cdot)$, the procedure is similar except we randomly divide the $n_k$ samples from study $k$ into $Q$ evenly-sized folds. We let $\hat{e}_k^{-q_k(i)}(\cdot)$ denote the estimate based on all $n_k$ samples from study $k$ except for those in $q_k(i)$, where $q_k(i) \in \{1, \ldots, Q\}$ denotes the index of the fold that individual $i$ of study $k$ belongs in. 
 We minimize $\hat{L}_n(\{\tau_k(\cdot)\}_{k=1}^K) + \Lambda_{\tau_k}$ with respect to $\{\tau_k(\cdot)\}_{k=1}^K$, where
\begin{equation}
\label{multi_study_R_loss}
   \hat{L}_n\left(\{\tau_k(\cdot)\}_{k=1}^K\right)= \frac{1}{n}\sum_{i=1}^n\left[\{Y_{i}-\hat{m}^{-q(i)}(X_{i})\}-\sum_{k=1}^K\{A_{i}-\hat{e}^{-q_k(i)}_k(X_{i})\}\hat{p}^{-q(i)}(k \mid X_i){\tau}_k(X_{i})\right]^2
\end{equation}

\noindent is the plug-in multi-study $R$-loss. In the multi-study $R$-learner framework, we estimate nuisance functions with cross-fitting in Steps 1 and 2 and substitute the estimates into the multi-study $R$-loss (\ref{multi_study_R_loss}) to estimate study-specific HTEs in Step 3. Because $\tau(x) = \sum_{k=1}^K \tau_k(x)p(k \mid x)$ by definition, we can obtain $\hat{\tau}(x)$ for individuals with covariates $x$ by calculating the weighted average of study-specific HTEs, i.e., $\hat{\tau}(x) = \sum_{k=1}^K \hat{\tau}_k(x)\hat{p}(k \mid x)$.

\section{Theoretical Analysis}
\label{s:theory}
The goal of our theoretical analysis is three-fold. First, we show that the difference between the oracle multi-study $R$-loss $L_n\left(\{\tau_k(\cdot)\}_{k=1}^K\right)$ in Eqn (\ref{oracle_loss}) and the plug-in version $\hat{L}_n\left(\{\tau_k(\cdot)\}_{k=1}^K\right)$ in Eqn (\ref{multi_study_R_loss}) diminishes at a relatively fast rate with $n$. Second, we show that the multi-study $R$-learner is asymptotically unbiased and normally distributed in the case of series estimation, a nonparametric regression method that approximates an unknown function based on $d$ basis functions, where $d$ is allowed to grow with $n$ to balance the trade-off between bias and variance \citep{wasserman2006all, belloni2015some}. Series estimation is widely-studied  and thus serves as an ideal case study for examining the asymptotic behavior of our approach. Third, we show that the multi-study $R$-learner is more efficient than the $R$-learner when there is between-study heterogeneity in the propensity score. Specifically, we compare the multi-study $R$-learner estimator $\hat{\tau}(\cdot)$ to the study-specific $R$-learner estimator $\hat{\tau}^{SS}(\cdot) = \sum_{k=1}^K \hat{\tau}^R_k(\cdot)\hat{p}(k\mid \cdot)$, where we fit an $R$-learner $\hat{\tau}_k^R(\cdot)$ on studies $k=1,\ldots, K$ separately and then ensemble the results. 

Under the series estimation framework, $\tau_k(x)$ can be approximated by a linear combination of some pre-specified basis functions $v_k(x) = (v_{k,1}(x), \ldots, v_{k,d_k}(x))^\intercal$. That is,
$
\tau_k(x) = v_k^\intercal(x)\beta_k + r_f,
$
where $\beta_k \in \mathbb{R}^{d_k}$ are the corresponding combination weights and $r_f$ the approximation error for function $f$. We allow $d_k$ to grow with the sample size $n_k$. The set of basis functions is not necessarily the same across studies; examples of these functions include polynomial and regression splines, among others. If the approximation error of $\tau_k(x)$ is asymptotically negligible, then the oracle multi-study $R$-loss in (\ref{oracle_loss}) can be expressed as a quadratic function of $\beta = (\beta_1^\intercal, \ldots, \beta_K^\intercal)^\intercal$,
$$
    L_n(\beta) = \frac{1}{n}\sum_{i=1}^n\left[\{Y_{i}-{m}(X_{i})\}-u_i^{\intercal}\beta\right]^2,
$$
where $u_i = u(A_i, X_i) = W(A_i, X_i)Z(X_i)v(X_i)$,
\begin{align*}
    W(A_i, X_i) = \text{blkdiag}[\{W_k(A_i, X_i)\}_{k=1}^K], &\quad W_k(A_i, X_i) = \{A_i - e_k(X_i)\} \mathbb I_{d_k},\\
    Z(X_i) = \text{blkdiag}[\{Z_k(X_i)\}_{k=1}^K], &\quad
    Z_k(X_i) = p(k \mid X_i) \mathbb I_{d_k},
\end{align*}
$v(x) = (v^{\intercal}_1(x), \ldots, v^{\intercal}_K(x))^\intercal$, and $\mathbb I_n$ is the $n$-dimensional identity matrix. The plug-in multi-study $R$-loss is
$$
    \hat{L}_n(\beta) = \frac{1}{n}\sum_{i=1}^n\left[\{Y_{i}-\hat{m}^{-q(i)}(X_{i})\}-\hat{u}_i^{\intercal}\beta\right]^2,
$$ and estimates of $\tau_k(\cdot)$, $k = 1,\ldots, K$, can be obtained by deriving the optimizer,
$$
    \hat{\beta} = \argmin_b \frac{1}{n}\sum_{i=1}^n\left[\{Y_{i}-\hat{m}^{-q(i)}(X_{i})\}-\hat{u}_i^{\intercal}b\right]^2.
$$ 

\subsection{Assumptions}
To formally show that the multi-study $R$-learner estimator $\hat{\tau}(x)$ is asymptotically unbiased and normal for any $x \in \mathcal{X}$, we make the following assumptions. We use the notation $a \lesssim b$ to denote $a \leq cb$ for some constant $c > 0$ that does not depend on $n$.

\begin{assumption}[Boundedness]\label{assumption4}
    $\norm{\tau_k(x)}_{\infty}$ and $E\left[\{Y - m(X)\}^2\mid X, A\right]$ are bounded, for any $X \in \mathcal{X}$, $A\in \{0,1\}$, $k = 1, \ldots, K$.
\end{assumption}

\begin{assumption}[Estimation accuracy]\label{assumption5} $E\left[\{m(X) - \hat{m}(X)\}^2\right] and \\
\indent E\left(\left[ \{e_k(X)-A\}p(k\mid X)-\{\hat{e}_k(X)-A\}\hat{p}(k \mid X) \right]^2\right)$ are $O(a_n^2)$, where $a_n$ is a sequence\\ \indent such that $a_n = O(n^{-r})$ with $r > 1/4$. 
\end{assumption}

\noindent The assumptions above are required to show that the difference between the oracle and plug-in multi-study $R$-loss diminishes at a relatively fast rate with $n$. To show asymptotic normality, we assume the membership probabilities follow a multinomial regression model, $p(k\mid x) = 
\frac{exp(\tilde{v}_k(x)^\intercal \gamma_k)}{1 + \sum_{k=1}^{K-1} exp(\tilde{v}_k(x)^\intercal \gamma_{k})}$ for $k = 1, 2, \ldots, K - 1$, where $\tilde{v}_k(x)$ are pre-specified basis functions for study $k$, and $\gamma_k$ is the corresponding parameter vector. Let $w_i = (y_i, x_i, a_i, s_i)$ denote the vector of observations for the $i$th individual. We are interested in the parameter vector $\theta = (\beta, \gamma)$, where $\gamma = \begin{bmatrix}\gamma_1^\intercal, \gamma_2^\intercal, \ldots, \gamma_{K-1}^\intercal\end{bmatrix}^\intercal$, that satisfies the system of equations,
$$\Psi_n(\theta) = \frac{1}{n}\sum_{i=1}^n \psi(w_i, \theta) = 0,$$ where $\psi(w_i, \theta) = \partial h(w_i, \theta)/\partial \theta = \partial \left[\{y_i-m(x_i)\} - u_i^\intercal \beta\right]^2/\partial \theta.$ We state the following regularity and identification assumptions.
\begin{assumption}[Regularity]\label{assumptionConsistency} Let $w$ be a vector taking values in $\mathcal{W}$. (a) For each $\theta \in \Theta$, $h(\cdot, \theta)$\\
\indent is a Borel measurable function on $\mathcal{W}$.\\
\indent (b) For each $w \in \mathcal{W}$, $h(w, \cdot)$ is a continuous function over the compact parameter space $\Theta$.\\
\indent (c) $|h(w, \theta)| \leq b(w)$ for all $\theta \in \Theta$, where $b$ is a nonnegative function such that $E[b(w)] < \infty.$
\end{assumption}

\begin{assumption}[Identification]\label{assumptionIdentify} $E[h(w, \theta_0)] < E[h(w, \theta)]$ for all $\theta \in \Theta$, $\theta \neq \theta_0.$
\end{assumption}

\noindent Assumptions~\ref{assumptionConsistency} and \ref{assumptionIdentify} allow us to establish consistency, i.e., $\hat{\theta} \overset{p}{\rightarrow} \theta_0,$ where $\hat{\theta} = \argmax_{\theta \in \Theta} \frac{1}{n} \sum_{i=1}^n h(w_i, \theta)$ and $\theta_0 = \argmax_{\theta \in \Theta} E[h(w, \theta)].$ Let $H(w_i, \theta) = \partial \psi(w_i, \theta)/\partial \theta$ denote the Hessian. We now state the additional assumptions needed for establishing asymptotic normality.

\begin{assumption}[Regularity]\label{assumptionAsymptoticNormality} (a) $\theta_0$ is an interior point of $\Theta$.\\
\indent (b) $\psi(w, \cdot)$ is continuously differentiable on the interior of $\Theta$ for all $w \in \mathcal{W}$.\\
\indent (c) Each element of $H(w, \theta)$ is bounded in absolute value by $b(w),$ where $E[b(w)] < \infty$.\\
\indent (d) $E[H(w, \theta_0)]$ is positive definite. \\
\indent (e) $E[\psi(w, \theta_0)\psi^\intercal(w, \theta_0)] < \infty$.
\end{assumption}



\begin{assumption}[Approximation]\label{assumption7}
(a) For each $n$ and $d$, there are finite constants $c_{d}$ and $l_{d}$ such \indent that for each $f \in \mathcal{G}$, $$\left\|r_{f}\right\|_{F, 2}=\sqrt{\sum_{a = 0}^1 \left\{\int_{x \in \mathcal{X}} r_{f}^{2}(a, x) d F(x)\right\}} \leq c_{d}$$ \indent and $\left\|r_{f}\right\|_{F, \infty}=\sup _{x \in \mathcal{X}, a \in \{0, 1\}}\left|r_{f}(a, x)\right| \leq l_{d} c_{d},$
where $\mathcal{G}$ is a function class. 

(b) $\sup_{x \in \mathcal{X}} E[\epsilon^2 I\{|\epsilon| > M\} \mid X = x] \rightarrow 0$ as $M \rightarrow \infty$ uniformly over $n$.

(c) Let $\underline{\sigma}^2 = \inf_{x \in \mathcal{X}} E[\epsilon^2 \mid X = x]$. We assume that 
$\underline{\sigma}^2 \gtrsim 1$.

(d) Let $\xi_d = \sup_{a,x} \norm{u(a,x)}$, we assume that $$\{\xi_{d}^{2}\log(d)/n\}^{1/2}\left(1+\sqrt{d} l_{d} c_{d}\right) \rightarrow 0~~\text{and}~~l_{d} c_{d} \rightarrow 0.$$

(e) $\lim\limits_{n\to\infty} \sqrt{n/d}\cdot l_d c_d =
 0.$


\end{assumption}

\begin{assumption}[Homoscedasticity]\label{assumption10} $E[\epsilon^2 \mid X, A, S = k] = \sigma^2_k$ and $E[\epsilon^2 \mid X, A] = \sigma^2.$
    
\end{assumption}

In Assumption~\ref{assumption7}a, the constants $c_d$ and $l_d$ together characterize the approximation properties of $u(a, x)^\intercal\beta$. Assumption~\ref{assumption7}b is a mild uniform integrability condition, and it holds if for some $m > 2, \sup_{x\in \mathcal{X}} E[|\epsilon|^m \mid X = x] \lesssim 1$. Assumption~\ref{assumption7}c is a mild regularity condition. Assumptions~\ref{assumption7}d and \ref{assumption7}e ensure that the impact of unknown design is negligible, and the approximation error is negligible relative to the estimation error, respectively. Assumption \ref{assumption10} posits homoscedastic errors. Under Assumptions \ref{assumption1} (consistency) and \ref{assumption2} (mean unconfoundedness within study), the homoscedasticity assumption entails that $\sigma^2 = Var(Y\mid X, A) = \sum_{k=1}^K \sigma^2_k p(k\mid X) + Var(m_S(X)\mid X)$ does not vary as a function of $X$. For example, the first term will not depend on $X$ if $\sigma_k$ is identical across studies. The second term is homoscedastic for certain formulations of $m_S(X),$ including but not limited to $m_S(X) = h_1(S) + h_2(X)$ (i.e., random study-specific intercept model), where $h_1(\cdot)$ and $h_2(\cdot)$ denote functions. 
\subsection{Results}
\begin{lemma}
\label{lem:1}
Under Assumptions~\ref{assumption4} and \ref{assumption5}, $\hat{L}_n(\beta) = L_n(\beta) + O_p(a^2_n)$.
\end{lemma}
\begin{theorem}[Asymptotic Normality]
\label{thm:1}
Let $g(\beta, \gamma) = \sum_{k=1}^K p(k\mid x)v^\intercal_k(x)\beta_k$. Under Assumptions~\ref{assumption1}-\ref{assumption10}, for any $x \in \mathcal{X} \subseteq \mathbb{R}^p,$ 
$$
\sqrt{n}\{\hat{\tau}(x) - \tau(x)\} \overset{d}{\rightarrow} N(0, (\mathcal{D}g)A^{-1}_0B_0A^{-1}_0(\mathcal{D}g)^\intercal),
$$
where $A_0 = E[H(\omega, \theta_0)],$ $B_0 = E[\psi(\omega, \theta_0) \psi^{\intercal}(\omega, \theta_0)]$, $\mathcal{D}g = [\partial^\intercal g(\beta,\gamma)/\partial \beta \hspace{0.3cm} \partial^\intercal g(\beta,\gamma)/\partial \gamma]^\intercal$. 
\end{theorem} 
Proofs are provided in the Supplementary Material. Lemma~\ref{lem:1} states that the difference between the oracle (\ref{oracle_loss}) and plug-in (\ref{multi_study_R_loss}) loss functions diminishes with rate $a^2_n$, which implies that $\hat{\beta} = \argmin_b n^{-1}\sum_{i=1}^n \{Y_i - m(X_i) - u_i^{\intercal}b\}^2 + O_p(a^2_n).$ Theorem~\ref{thm:1} provides the asymptotic normality result for the multi-study $R$-learner estimator $\hat{\tau}(x)$ for any $x \in \mathcal{X}$. To examine efficiency gains, we compare the multi-study $R$-learner to the study-specific $R$-learner estimator where we fit an $R$-learner on each study separately and then ensemble the results,
\begin{equation}
\label{eqn:esrlearner}
    \hat{\tau}^{SS}(x) = \sum_{k=1}^K \hat{p}(k\mid x)v_k^\intercal(x)\hat{\beta}^R_k,
\end{equation}
\noindent where for $\hat{u}^R_{k,i} =\hat{p}(k\mid X_i)^{-1}\hat{u}_{k,i}, \hat{\beta}^R_k = \argmin_b {n_k}^{-1}\sum_{i=1}^{n_k} \left[\{Y_{i}-\hat{m}_k^{-q_k(i)}(X_{i})\}-(\hat{u}^R_{k,i})^{\intercal}b\right]^2.$
\begin{theorem}[Efficiency]
\label{thm:2}
    Suppose Assumptions~\ref{assumption1}-\ref{assumption10} hold. When $K=2$, if for all $x \in \mathcal{X},$ $p(1\mid x) = p(2\mid x)$, $v_1(x) = v_2(x)$, $e_1(x)\{1 - e_1(x)\} = ce_2(x)\{1-e_2(x)\}$ for $c\in \mathbb{R}^+$ such that $c\neq 1$, and $\sigma^2 > (1 + c)(2\sqrt{c})^{-1}$, then the multi-study $R$-learner is more efficient than the study-specific $R$-learner, i.e., $var(\hat{\tau}(x))_{\text{Asy}} < var(\hat{\tau}^{SS}(x))_{\text{Asy}}.$
\end{theorem}
\noindent A proof is provided in the Supplementary Material. Theorem~\ref{thm:2} states that in the two-study setting, the multi-study $R$-learner is more efficient than the study-specific $R$-learner when there is between-study heterogeneity in the propensity score models (i.e., $e_1(\cdot)\{1 - e_1(\cdot)\} = ce_2(\cdot)\{1 - e_2(\cdot)\}$ for some $c\neq 1$) and the conditional variance of the error term $\sigma^2$ is greater than the threshold $(1 + c)(2\sqrt{c})^{-1}$. In practice, an example for which $e_1(\cdot)\{1 - e_1(\cdot)\} = ce_2(\cdot)\{1 - e_2(\cdot)\}$ holds is when the randomization ratio differs between studies. An exception is when the randomization ratio in one study is the reciprocal of that in the other, i.e., 1:3 v.s. 3:1, as this corresponds to the $c = 1$ setting in Theorem \ref{thm:2}. Under Assumption \ref{assumption10}, note that $\sigma^2 = Var(Y\mid X, A) = \sum_{k=1}^K \sigma^2_k p(k\mid X) + Var(m_S(X)\mid X)$ captures between-study heterogeneity in the mean outcome model $m_S(X)$ across studies $S = 1, \ldots, K.$ As such, the threshold $(1 + c)(2\sqrt{c})^{-1}$ can be thought of as a between-study-heterogeneity transition point beyond which the multi-study $R$-learner is more efficient. 
\section{Ovarian Cancer Simulations}
\label{s:simulations}
We conducted extensive simulations comparing the performance of the multi-study $R$-learner to the study-specific $R$-learner in Eqn (\ref{eqn:esrlearner}). Specifically, we considered four scenarios (A-D) to explore combinations of different functional forms of $\tau_k(\cdot)$ (linear vs.\ non-linear) and study designs (randomized vs.\ observational). For each scenario, we generated $K = 4$ studies with $n = 600$ samples. We randomly sampled $p = 40$ gene expression covariates from the \texttt{curatedOvarianData} R package (\cite{ganzfried2013curatedovariandata}) to reflect realistic and potentially heterogeneous covariate distributions. For study $k$, we generated study-specific HTEs, $\tau_k(\tilde{X}_i) = \tilde{X}_i\beta_k + Z_i \eta_k$, where $\tilde{X}_i$ was the basis-expanded predictor vector, $\beta_k \sim \mbox{MVN}(2, \mathbb{I}_{d_k})$, and $Z_i \in \mathbb{R}^{q}$ was a subset of $\tilde{X}_i$ that corresponded to the random effects $\eta_k \in \mathbb{R}^q.$ The random effects have $E[\eta_k] = 0$ and $Cov(\eta_k) = \text{diag}(\sigma^2_{1}, \ldots, \sigma^2_{q}).$ If $\sigma^2_{j} > 0$, then the effect of the $j$th basis-expanded covariate varies across studies; if $\sigma^2_j = 0,$ then the covariate has the same effect in each study. 

In Scenarios A and B, we assumed that $\tau_k(\cdot)$s were linear; in Scenarios C and D, we assumed non-linear $\tau_k(\cdot)$s, where the first coordinate in $X_{i}$ was expanded with a cubic spline at knot = 0. 
We assigned each observation to one of $K = 4$ studies based on its membership probability 
$p(k \mid X_i) \sim \text{Multinom}\left(n=600, K = 4, p_k = exp(X_i \gamma_k)\{1 + \sum_{k=1}^{K-1} exp(X_i \gamma_k)\}^{-1} \text{ for } k = 2, 3, 4\right)$ where $\gamma_k \sim \text{MVN}(0, I)$. We generated the mean counterfactual model under treatment $a = 0$ as $\muko(X_i) = X_i \nu^{(0)}_k$ with $\nu^{(0)}_k \sim \text{MVN}(0, I)$. In Scenarios A and C, the studies were randomized, so $e_k(x) = 0.5$ for all $x$. In B and D, the studies were observational, so we generated $e_k(X_i) = \text{expit}(X_i \delta_k)$ where $\delta_k \sim MVN(0, I)$. We estimated the nuisance functions $m_k(\cdot), e_k(\cdot),$ $k = 1, \ldots K$ and membership probabilities $p(k\mid \cdot)$ with elastic net. 

In each scenario, we introduced between-study heterogeneity by varying the magnitude of the regression coefficients through $\sigma^2_{\tau}$. To simulate between-study heterogeneity in confounding, we set the proportion of overlap in covariates' support to 0.5 so that different studies had different confounders. For each simulation replicate, we randomly divided each study into training and testing with a 70/30 split. All analyses were performed using R version 4.0.2.

\begin{figure}
    \includegraphics[scale=0.33]{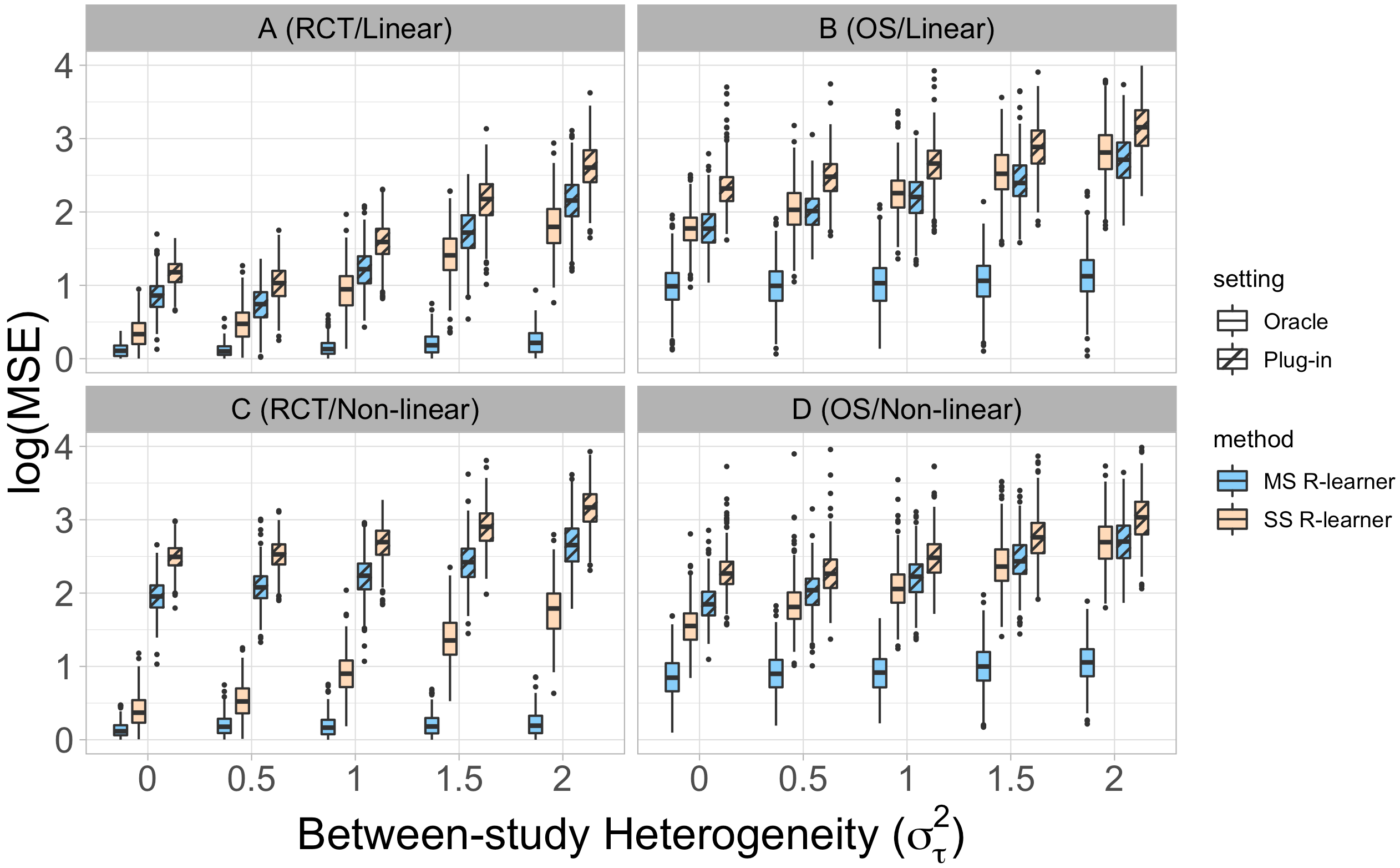}
	     \caption{Log mean squared error (MSE) of the multi-study $R$-learner (MS $R$-learner) and study-specific $R$-learner (SS $R$-learner) across different levels of between-study heterogeneity $\sigma^2_{\tau}$.}
      \textit{\small RCT = randomized clinical trial; OS = observational study; Linear = linear heterogeneous treatment effect; Non-linear = non-linear heterogeneous treatment effect}
      \label{fig:1}
\end{figure}

Fig. \ref{fig:1} shows the log mean squared error (MSE) of the multi-study $R$-learner and study-specific $R$-learner on the unseen test set. In each scenario, we considered two optimization settings: the oracle setting where the true nuisance functions are known and the plug-in setting where the nuisance functions need to be estimated from data. Overall, the multi-study $R$-learner generally outperformed the study-specific $R$-learner on the test set across all levels of between-study heterogeneity. This trend was observed in both the oracle and plug-in settings. In some cases, i.e., Scenario B and at higher levels of $\sigma^2_{\tau}$ in Scenario D, the plug-in multi-study $R$-learner performed similarly to the oracle study-specific $R$-learner, suggesting that the multi-study $R$-learner with estimated nuisance functions was comparable to the study-specific $R$-learner with known nuisance functions.
\section{Breast Cancer Data Application}
\label{s:data_app}
To illustrate the multi-study $R$-learner in a realistic application involving an RCT and observational study, we used data from the \texttt{curatedBreastData} R package (\cite{planey2015package}). In contrast to the previous data experiment, we now let both the baseline signal and the propensity scores come from real data. Our goal was to estimate HTEs of neoadjuvant chemotherapy on breast cancer patients. In practice, neoadjuvant chemotherapy is commonly used to reduce the size of breast cancer before the main treatment, e.g., surgery. Because different sub-types of breast cancer behave and proliferate in different ways \citep{hwang2019impact},  it's important to characterize the treatment heterogeneity of neoadjuvant chemotherapy as individuals who don't respond run the additional risk of delaying surgery. Thus, the purpose of our data experiment was to characterize the HTE of anthracyline ($A$) versus taxane ($T$), two neoadjuvant chemotherapy regimens for early breast cancer.

The outcome of interest was pathological complete response, defined as disappearance of all invasive cancer in the breast after completion of neoadjuvant chemotherapy ($Y=1$) or otherwise ($Y=0$). We identified $K = 2$ studies where patients were administered the neoadjuvant chemotherapy of interest. The first study (GSE21997) was a randomized trial of $n_1 = 94$ women aged between 18 and 79 with stage II-III breast cancer (\cite{martin2011genomic}). The second study (GSE25065) was an observational study of $n_2 = 168$ women who were HER2 negative (HER2-) with stage I-III breast cancer (\cite{hatzis2011genomic}). We focused on five clinical covariates (age, histology grade, HR+, PR+, and HER2+ status) and eight genes (\textit{SCUBE2, MMP11, BCL2, MYBL2, CCNB1, ACTB, TFRC, GSTM1}) from Oncotype DX, a gene assay used by oncologists that predicts whether a patient would benefit from chemotherapy (\cite{sparano2008development}).

A challenge with illustrating HTE estimators on real data is that we do not observe both potential outcomes. Therefore, we simulated study-specific treatment effects to make the task of estimating HTEs non-trivial. For $k \in \{1, 2\}$, $x \in \mathcal{X}$, and $ i \in \{1,\ldots, n\}$, we generated
    $\tau_k(x) = \text{pr}(Y_i(1) \mid X_i=x, S_i=k) - \text{pr}(Y_i(0) \mid X_i=x, S_i=k) =  p^{(1)}_{k}(x) -  p^{(0)}_{k}(x),$
where the potential outcome probabilities $p^{(1)}_{k}(x)$ and $p^{(0)}_{k}(x)$ were linear functions of the five clinical features and eight Oncotype DX genes on the logit scale. Because all participants in study~2 were HER2-, we generated their potential outcome probabilities such that the HER2 effect is the same for all participants in that study. 
To this end, HER2 status was not a confounder in study 2. To introduce between-study heterogeneity, we assigned random effects with mean 0 and variance $\sigma^2_{\tau}$ to the five clinical features (age, histology grade, HR+, PR+, HER2+) and eight genes (\textit{SCUBE2, MMP11, BCL2, MYBL2, CCNB1, ACTB, TFRC, GSTM1}). For individuals in study $k$ with covariate $x$, we generated counterfactual outcomes $Y(a)$ from a Bernoulli distribution with probability $p^{(a)}_{k}(x)$ and set $Y = Y(a)$. We randomly divided the data into a training ($n_{\text{train}}= 232$) and test set ($n_{\text{test}} = 30$). 

In addition to the study-specific $R$-learner, we compared our method to \cite{wu2021integrative}'s integrative $R$-learner, which was designed to leverage data from an RCT and observational study. We estimated the nuisance functions from the training data using Lasso with tuning parameters selected by cross-validation. Because study 1 was an RCT, we used 0.5 as the propensity score. We estimated $\hat{e}_2(\cdot)$ using training data from study 2. To estimate the membership probabilities for the multi-study $R$-learner, we fit a logistic regression model to the training data's study labels, $S \in \{1, 2\}$. Next, we optimized the loss functions to estimate the HTEs and calculated the MSE on the test set for each approach. For the integrative $R$-learner, we used the penalty scale searching procedure to select optimal shrinkage parameters (c.f. Algorithm 1 in \cite{wu2021integrative}). We applied the transformation $2 \times \text{expit}(\hat{\tau}(\cdot))-1$ to ensure that $\hat{\tau}(\cdot)$ was between -1 and 1. This transformation is intuitive in that positive values of $\hat{\tau}(\cdot)$ correspond to treatment $A$ being more beneficial than $T$, and vice-versa for negative values.

\begin{figure}
\centering
\includegraphics[scale=0.45]{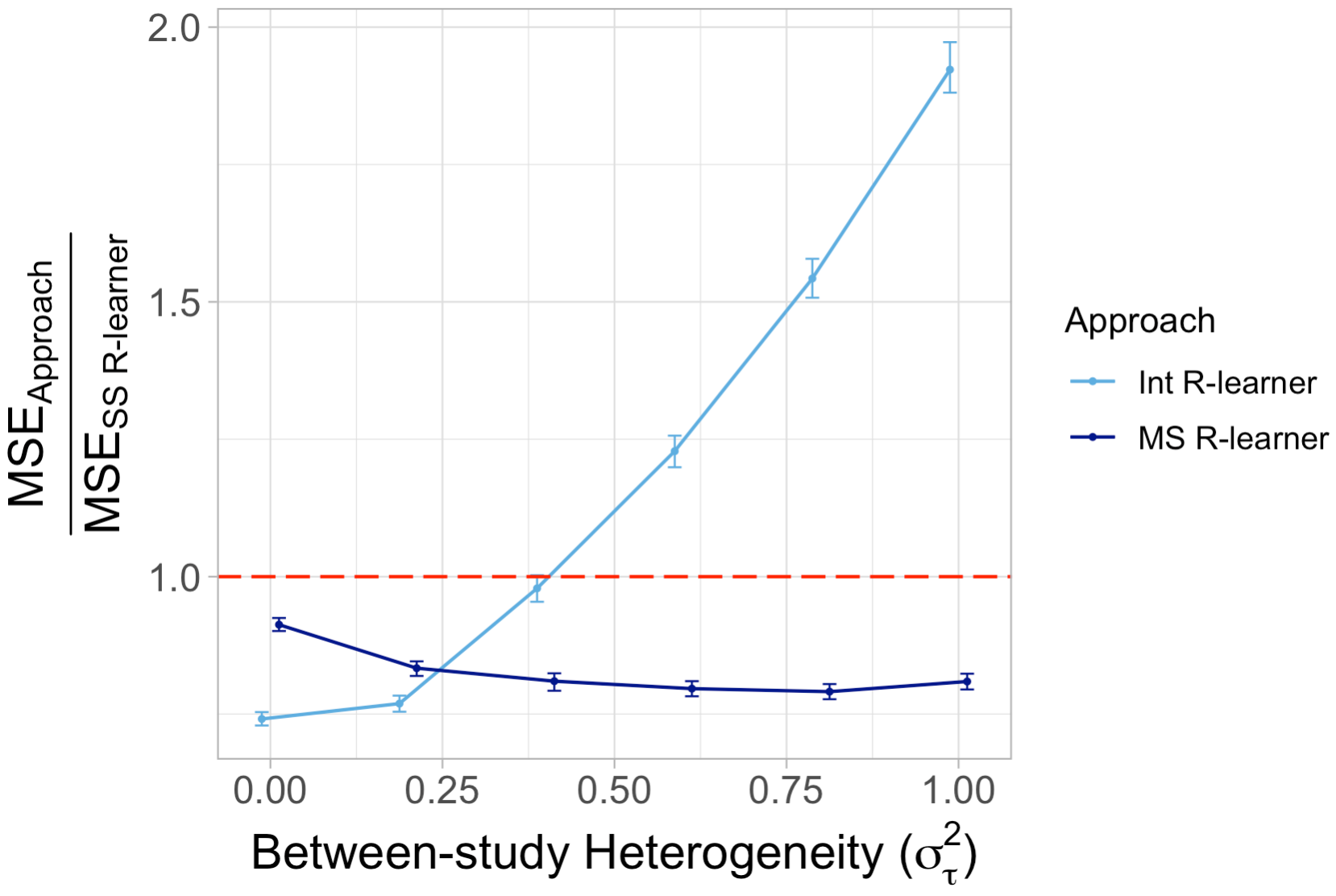}
    \caption{Mean squared error (MSE) ratio comparing the integrative $R$-learner (Int $R$-learner; light blue)  and the multi-study $R$-learner (MS $R$-learner; dark blue) to the study-specific $R$-learner (SS $R$-learner; reference) across various levels of between-study heterogeneity $\sigma^2_{\tau}$. Solid circles and vertical bars represent the average performance ratios and 95\% confidence intervals, respectively.}
    
    \label{fig:2}
\end{figure}

Fig. \ref{fig:2} shows the log MSE ratio comparing the performance of the three approaches. When there is no between-study heterogeneity $(\sigma_{\tau} = 0)$, both the integrative and multi-study $R$-learner outperformed the study-specific version. As between-study heterogeneity increased (i.e., further deviation from the transportability assumption), the multi-study $R$-learner showed favorable performance.

\section{Discussion}
\label{s:discussion}
We proposed the multi-study $R$-learner for estimating HTEs across multiple studies. By linking cross-study estimates of nuisance functions and HTEs via membership probabilities, the approach borrows information across potentially heterogeneous studies. Our setup requires that practitioners have access to participant-level data from multiple studies measuring a common set of covariates. Recent facilitation of systematic data sharing has led to increased access to standardized data across multiple sites. Prominent examples include the U.S. National Institutes of Health's \textit{All of Us} Research Program and the UK Biobank, which are national initiatives that standardize data across multiple collection sites by employing uniform protocols for sample gathering, health assessments, and data entry \citep{all2019all, biobank2014uk}. 

A related line of work is federated learning across distributed data sites. In this setting, information exchange between sites may be restricted due to privacy or feasibility considerations, prohibiting pooled analyses \citep{maro2009design, mcmahan2017communication}. As such, study sites leverage models or parameters derived from other sites without sharing individual-level data. In the context of HTE estimation, \cite{tan2022tree} proposed a tree-based ensemble approach that combines models across data sites. \cite{vo2022adaptive} performed federated causal inference through adaptive kernel functions on observational studies. Similar to this line of work, the multi-study $R$-learner performs cross-site learning by computing study-specific nuisance functions and HTEs for all individuals. Currently, the multi-study $R$-learner loss function requires centralized access to individual-level data from all sites, and, as a result, is not directly applicable to distributed data. Thus, we leave this extension to future work.

The proposed framework can be generalized to incorporate different multi-study learning strategies for estimating $m(\cdot)$. Generally, when studies are homogeneous, \cite{patil2018training} showed that merging all studies and training a single model can lead to improved accuracy due to increased sample size; as between-study heterogeneity increases, multi-study ensembling is preferred. The empirical and theoretical trade-offs between merging and multi-study ensembling have been explored in detail for ML techniques, including linear regression (\cite{guan2019merging}), random forest (\cite{ramchandran2020tree}), gradient boosting (\cite{shyr2022multi}), and multi-study stacking \citep{ren2020cross}. Because estimation of $m(\cdot)$ directly impacts the downstream analysis of $\tau(\cdot)$, exploring the empirical and theoretical implications of these strategies in the context of the multi-study $R$-learner is an interesting avenue for future work.


\backmatter



\section*{Funding}
Research reported in this publication was supported by the U.S. National Institutes of Health National Library of Medicine under Award Number 1K99LM014429-01 (C.S.), National Cancer Institute under Award Number 5T32CA009337-40 (C.S.) and the National Science Foundation under Award Number NSF-DMS 2113707 (G.P. and B.R.).

\section*{Data Availability}
Data used in this study can be found in the R packages \texttt{curatedOvarianData} and\\
\texttt{curatedBreastData} at https://www.bioconductor.org. Code to reproduce results from this paper can be found at https://github.com/cathyshyr/multi-study-r-learner.



\bibliographystyle{biom}
\bibliography{biomsample}
\section{Figures}

\newpage
\clearpage
\newpage

\section{Appendix}
\begin{proof}{(Proposition 1)}
\begin{align*}
    E[\epsilon_i \mid X_i, A_i] =& 
    E[Y_i(A_i) \mid A_i, X_i] - E\left[\left.\sum_{k=1}^K \left\{\muko(X_i) + A_i\tau_k(X_i)\right\}p(k \mid X_i)\right|A_i, X_i\right]\\
    = &E[Y_i(A_i) \mid A_i, X_i] - \sum_{k=1}^K p(k \mid X_i)E[\underbrace{E[Y_i(0) \mid X_i, S_i = k]}_{\muko(X_i)} \mid A_i, X_i]\\
     &- A_i \sum_{k=1}^K p(k\mid X_i) E[\underbrace{E[Y_i(1) - Y_i(0) \mid X_i, S_i = k]}_{\tau_k(X_i)} \mid A_i, X_i]\\
    = &E[Y_i(A_i) \mid A_i, X_i] - (1-A_i)\sum_{k=1}^K p(k \mid X_i)E[E[Y_i(0) \mid X_i, S_i = k] \mid A_i, X_i]\\
    &- A_i\sum_{k=1}^K p(k \mid X_i)E[E[Y_i(1)\mid X_i, S_i = k] \mid A_i, X_i]\\
    =&E[Y_i(A_i) \mid A_i, X_i] -(1-A_i)\sum_{k=1}^K p(k \mid X_i)E[E[Y_i(0) \mid A_i=0,X_i, S_i = k] \mid A_i, X_i]\\
    &- A_i\sum_{k=1}^K p(k \mid X_i)E[E[Y_i(1)\mid A_i = 1,X_i, S_i = k]\mid 
    A_i, X_i]\\
    = &0
\end{align*}
The second to last equality holds by Assumption 2. The last equality holds because for $A_i = a \in \{0, 1\}$, we have

\begin{align*}
    &E[Y_i(a) \mid A_i = a, X_i] - \sum_{k=1}^K p(k \mid X_i)E[E[Y_i(a)\mid A_i = a,X_i, S_i = k] \mid A_i = a, X_i]\\
    = & E[Y_i(a)\mid A_i = a, X_i] - \sum_{k=1}^K p(k \mid X_i)E[Y_i(a)\mid A_i = a,X_i]\\
    =& E[Y_i(a) \mid A_i = a, X_i] - E[Y_i(a) \mid A_i = a, X_i]\\
    =&0.
\end{align*}
\end{proof}
\clearpage

\begin{proof}(Lemma 1)
Recall that $$ \hat{L}_n(\beta) = \frac{1}{n}\sum_{i=1}^n\left[\{Y_{i}-\hat{m}^{-q(i)}(X_{i})\}-\hat{u}_i^{\intercal}\beta\right]^2$$
and $$ L_n(\beta) = \frac{1}{n}\sum_{i=1}^n\left[\{Y_{i}-{m}(X_{i})\}-u_i^{\intercal}\beta\right]^2.$$
We can re-write $\hat{L}_n(\beta)$ as 
$$ \hat{L}_n(\beta) = \frac{1}{n}\sum_{i=1}^n\left[\{Y_{i}-\hat{m}^{-q(i)}(X_{i})\}-\sum_{k=1}^K \{A_i - \hat{e}^{-q_k(i)}_k(X_i)\}\hat{p}^{-q(i)}(k\mid X_i)v_k(X_i)^\intercal \beta_k\right]^2,$$
where $\beta_k \in \mathbb{R}^{d_k}$ is the vector of coefficients for study $k$. Similarly, we can re-write $L_n(\beta)$ as 
$$ L_n(\beta) = \frac{1}{n}\sum_{i=1}^n\left[\{Y_{i}-m(X_{i})\}-\sum_{k=1}^K \{A_i - e_k(X_i)\}p(k\mid X_i)v_k(X_i)^\intercal \beta_k\right]^2.$$

\noindent We define the following notation:
\begin{align*}
    A_{m,i} &= m(X_i) - \hat{m}^{-q(i)}(X_i)\\
    A_{e_k,i} &= (e_k(X_i)-A_i)p(k\mid X_i) - (\hat{e}^{-q_k(i)}_k(X_i)-A_i)\hat{p}^{-q(i)}(k \mid X_i) \hspace{3em} k = 1, \ldots, K\\
    B_{m, i} &= Y_i - m(X_i)\\
    B_{e_k, i} &= (A_i - e_k(X_i))p(k \mid X_i) \hspace{3em} k = 1, \ldots, K\\
    g_k(X_i, \beta_k) &= v_k(X_i)^\intercal\beta_k\\
    A_{e,i} &= \sum_{k=1}^K A_{e_k,i}g_k(X_i; \beta_k)\\
    B_{e,i} &= \sum_{k=1}^K B_{e_k,i}g_k(X_i;\beta_k),
\end{align*}

\noindent By algebra, we have
\begin{align*}
    \hat{L}_n(\beta) =& \frac{1}{n}\sum_{i=1}^n \left[B_{m, i} + A_{m, i} - \sum_{k=1}^K (B_{e_k,i} + A_{e_k, i})g_k(X_i;\beta_k)\right]^2\\
    =& \frac{1}{n}\sum_{i=1}^n \left[B_{m, i} + A_{m, i} - B_{e,i}-A_{e,i}\right]^2\\
    =& L_n(\beta) + \frac{1}{n}\sum_{i=1}^n \left[A_{m,i}-A_{e,i}\right]^2+\frac{2}{n}\sum_{i=1}^n (B_{m,i}-B_{e,i})(A_{m,i}-A_{e,i})\\
    =& L_n(\beta) + \frac{1}{n}\sum_{i=1}^n A^2_{m,i} + \frac{1}{n}\sum_{i=1}^n A_{e,i}^2 - \frac{2}{n}\sum_{i=1}^n A_{m,i}A_{e,i} \\
    &+ \frac{2}{n}\sum_{i=1}^n B_{m,i}A_{m,i} -  \frac{2}{n}\sum_{i=1}^n B_{m,i}A_{e,i} -  \frac{2}{n}\sum_{i=1}^n B_{e,i}A_{m,i} +  \frac{2}{n}\sum_{i=1}^n B_{e,i}A_{e,i}
\end{align*}

\noindent \textbf{First term: $\frac{1}{n} \sum_{i=1}^n A^2_{m,i} = \frac{1}{n} \sum_{i=1}^n (m(X_i) - \hat{m}^{-q(i)}(X_i))^2$}\\

\noindent By Markov's inequality and Assumption 5, $\frac{1}{n}\sum_{i=1}^n A^2_{m,i}$ is $O_p(a^2_n).$\\

\noindent \textbf{Second term: $\frac{1}{n} \sum_{i=1}^n A^2_{e,i} = \frac{1}{n}\sum_{i=1}^n \left[\sum_{k=1}^K A_{e_k,i}g_k(X_i;\beta_k)\right]^2$}\\

\noindent We have
\begin{scriptsize}
\begin{align*}
     \frac{1}{n}\sum_{i=1}^n A^2_{e,i} &= \sum_{k=1}^K \left[\frac{1}{n} \sum_{i=1}^n \left\{(e_k(X_i) - A_i)p(k\mid X_i) - (\hat{e}^{-q_k(i)}_k(X_i) - A_i)\hat{p}^{-q(i)}(k \mid X_i)\right\}^2g_k(X_i;\beta_k)^2\right] \\
     &\quad + \sum_{k\neq k'} \left[\frac{2}{n}\sum_{i=1}^n \left\{(e_k(X_i) - A_i)p(k \mid X_i) - (\hat{e}^{-q_k(i)}_k(X_i)-A_i)\hat{p}^{-q(i)}(k \mid X_i)\right\}g_k(X_i;\beta_k)\right.\\
     &\hspace{6.3em} \left.\times \left\{(e_{k'}(X_i)-A_i)p(k'\mid X_i)-(\hat{e}^{-q_k(i)}_{k'}(X_i)-A_i)\hat{p}^{-q(i)}(k' \mid X_i)\right\}g_{k'}(X_i;\beta_{k'})\right]
\end{align*}
\end{scriptsize}

\noindent By Markov's inequality and Assumption 5, the first sum is $O_p(a^2_n).$ Similarly, the cross term is also $O_p(a^2_n).$ Therefore, we can conclude that $\frac{1}{n}\sum_{i=1}^n A^2_{e, i}$ is $O_p(a^2_n).$\\

\noindent \textbf{Third term: $\frac{1}{n} \sum_{i=1}^n A_{m,i}A_{e,i}$}\\

\noindent We have
\begin{tiny}
\begin{align*}
   &\frac{1}{n}\sum_{i=1}^n A_{m,i}A_{e,i}\\
   &=  \frac{1}{n}\sum_{i=1}^n \left\{(m(X_i) - \hat{m}^{-q(i)}(X_i))\left[\sum_{k=1}^K \left\{(e_k(X_i)-A_i)p(k\mid X_i) -(\hat{e}^{-q_k(i)}_k(X_i)-A_i)\hat{p}^{-q(i)}(k \mid X_i)\right\}g_k(X_i;\beta_k)\right]\right\}\\
   &\leq \frac{C_1}{n}\sum_{i=1}^n \left\{(m(X_i) - \hat{m}^{-q(i)}(X_i))\left[\sum_{k=1}^K (e_k(X_i)-A_i)p(k \mid X_i)-(\hat{e}^{-q_k(i)}_k(X_i)-A_i)\hat{p}^{-q(i)}(k \mid X_i)\right]\right\}\\
   &\leq C_1 \sqrt{\left(\frac{1}{n}\sum_{i=1}^n \left\{m(X_i) - \hat{m}^{-q(i)}(X_i)\right\}^2\right)\left\{\frac{1}{n}\sum_{i=1}^n\left[\sum_{k=1}^K (e_k(X_i) - A_i)p(k\mid X_i)-(\hat{e}^{-q_k(i)}_k(X_i)-A_i)\hat{p}^{-q(i)}(k \mid X_i)\right]^2\right\}}\\
   &= O_p(a^2_n)
\end{align*}
\end{tiny} 

\noindent for some constant $C_1$. The second line holds by Assumption 4, and the third line holds by Cauchy-Schwarz inequality. \\

\noindent \textbf{Fourth term: $\frac{1}{n}\sum_{i=1}^n B_{m,i}A_{m,i} = \frac{1}{n}\sum_{i=1}^n (Y_i - m(X_i))(m(X_i) - \hat{m}^{-q(i)}(X_i))$}\\

\noindent We define $$B^q_{mm} = \frac{1}{|\{i:q(i)=q\}|} \sum_{i:q(i)=q}B_{m,i}A_{m,i}$$

\noindent to be the sample average of $B_{m,i}A_{m,i}$ in the $q$th cross-fitting fold. By the triangle inequality, $$\left|\frac{1}{n}\sum_{i=1}^n (Y_i - m(X_i))(m(X_i) - \hat{m}^{-q(i)}(X_i))\right| \leq \sum_{q=1}^Q |B^q_{mm}|.$$
Therefore, it suffices to show that $B^q_{mm} = O_p(a^2_n).$ Let $\mathcal{I}^{-q} = \{X_i, A_i, Y_i, S_i: q(i) \neq q\}$ denote the set of observations that do not belong to the same data fold as observation $i.$ $B^q_{mm}$'s expectation is
\begin{align*}
    E\left(B^q_{mm}\right) &= E(B_{m,i}A_{m,i})\\
    &= E\left(E\left[B_{m,i}A_{m,i}\mid \mathcal{I}^{-q}, X_i\right]\right)\\
    &= E(A_{m, i}E\left[B_{m,i}\mid \mathcal{I}^{-q}, X_i\right])\\
    &= 0,
\end{align*}
where the last line follows from Assumption 1 and Assumption 2.\\

\noindent Next, its variance is
\begin{align*}
Var\left(B_{mm}^{q}\right) &=E\left\{\left(B_{mm}^{q}\right)^{2}\right\}\\
&=\frac{E\left\{\sum_{i: q(i)=q} B_{m, i}^{2} A_{m,i}^{2}+\sum_{i \neq j: q(i)=q, q(j)=q} B_{m, i} B_{m, j} A_{m, i} A_{m, j}\right\}}{|\{i: q(i)=q\}|^{2}} \\
&=\frac{E\left(B_{m,i}^{2} A_{m,i}^{2}\right)}{|\{i: q(i)=q\}|}+\frac{\sum_{i \neq j: q(i)=q, q(j)=q} E\left(B_{m, i} B_{m, j} A_{m, i} A_{m, j}\right)}{|\{i: q(i)=q\}|^{2}}\\
\end{align*}

\noindent For the first term, we have

\begin{align*}
    E\left(B_{m,i}^{2} A_{m,i}^{2}\right) &= E\left(E\left[B_{m,i}^2A_{m,i}^2\mid \mathcal{I}^{-q}, X_i\right]\right)\\
    &=E\left(A_{m,i}^2E\left[B_{m,i}^2\mid \mathcal{I}^{-q}, X_i\right]\right)\\ 
    &\leq C_2 E(A_{m,i}^2)\\
    &=O(a^2_n)
\end{align*}
for some constant $C_2.$ The second to last line holds from Assumption 4. And the last line holds from Assumption 5.\\

\noindent For the second term, we have
\begin{align*}
    E\left(B_{m, i} B_{m, j} A_{m, i} A_{m, j}\right) &=  E\left[E\left(B_{m, i} B_{m, j} A_{m, i} A_{m, j}\mid \mathcal{I}^{-q}, X_i\right)\right]\\
    &= E\left[A_{m,i}A_{m,j}E\left(B_{m, i} B_{m, j}\mid \mathcal{I}^{-q}, X_i\right)\right]\\
    &= E\left[A_{m,i}A_{m,j}E\left(B_{m,j} \mid \mathcal{I}^{-q}, X_i\right)E\left(B_{m, i} \mid \mathcal{I}^{-q}, X_i\right)\right]\\
    &= 0
\end{align*}
The second to last line follows because $B_{m,i}$ is independent of $B_{m,j}$ for $i \neq j$. The last line follows by the definition of $B_{m,i}$. Therefore, we have that 
$$Var(B^q_{mm}) = \frac{Q}{n}O(a^2_n) = O({a^2_n}/n),$$
where the first equality holds if the $Q$ folds have equal number of observations (i.e., $n/Q$ for each fold). Then by Chebychev's inequality, $\frac{1}{n}\sum_{i=1}^n B_{m,i}A_{m,i}=O_p(a^2_n/n).$\\

\noindent \textbf{Fifth and Sixth terms} can be shown to be $O_p(a_n^2/n)$ by a similar argument.\\  

\noindent \textbf{Seventh term:} 
\begin{small}
\begin{align*}
   \frac{1}{n}\sum_{i=1}^n B_{e,i}A_{e,i} &\leq \frac{C_4}{n}\sum_{i=1}^n A_{e,i}\\
   &= \frac{C_4}{\sqrt{n}} \sqrt{\frac{1}{n}\left(\sum_{i=1}^n A_{e,i}\right)^2}\\
  &= \frac{C_4}{\sqrt{n}} \sqrt{\frac{1}{n} \left(\sum_{i=1}^n A_{e,i}^2 + \sum_{i\neq j} A_{e,i}A_{e,j} \right)}\\
  &\leq \frac{C_5}{\sqrt{n}} \sqrt{\frac{1}{n}\sum_{i=1}^n A_{e,i}^2}\\
    &\leq \frac{C_6}{\sqrt{n}} \sqrt{\left(\frac{1}{n}\sum_{i=1}^n A_{e,i}^2\right)\left(\frac{1}{n}\sum_{i=1}^n A_{e,i}^2\right)}\\
  &= O_p(a^2_n/\sqrt{n})
\end{align*}
\end{small}

\noindent for some constants $C_4$, $C_5$, and $C_6$. The first line holds by Assumption 4, and the last line holds by Assumption 5.\\

\noindent Putting all seven terms together, $\hat{L}_n(\beta) - L_n(\beta)$ is dominated by $O_p(a^2_n)$ terms, so $\hat{L}_n(\beta) - L_n(\beta) = O_p(a^2_n).$
\end{proof}
\vspace{1em}

\begin{proof}(Theorem 1)

Under Assumptions 1-10, Lemma 1, and Lemma 4.1 (Pointwise Linearization) in \cite{belloni2015some}, we have that
$$\sqrt{n}(\hat{\theta} - \theta_0) = \sqrt{n}\left(\begin{bmatrix}
    \hat{\beta} - \beta\\
    \hat{\gamma} - \gamma
\end{bmatrix}\right) = \sqrt{n}\left(\begin{bmatrix}
\mathbb{G}_n[u_i\epsilon_i] + o_p(1) + O_p(\sqrt{n}a^2_n)\\
\hat{\gamma} - \gamma
\end{bmatrix}\right),$$
where $\mathbb{G}_n[f(w_i)] = \frac{1}{\sqrt{n}} \sum_{i=1}^n (f(w_i)-E[f(w_i)]).$ Under Assumption 5, $O_p(\sqrt{n}a^2_n)$ is negligible compared to $o_p(1)$, so we have 
$$\sqrt{n}(\hat{\theta} - \theta_0) = \sqrt{n}\left(\begin{bmatrix}
    \hat{\beta} - \beta\\
    \hat{\gamma} - \gamma
\end{bmatrix}\right) = \sqrt{n}\left(\begin{bmatrix}
\mathbb{G}_n[u_i\epsilon_i] + o_p(1)\\
\hat{\gamma} - \gamma
\end{bmatrix}\right).$$
\noindent By Theorem 4.2 (Pointwise Normality) in \cite{belloni2015some}, we conclude that
\begin{align*}
    \sqrt{n}(\hat{\theta} - \theta_0) = \sqrt{n}\left(\begin{bmatrix}
    \hat{\beta} - \beta\\
    \hat{\gamma} - \gamma
\end{bmatrix}\right) \overset{d}{\rightarrow} N(0, A_0^{-1}B_0A^{-1}_0) 
\end{align*}
where $B_0 = E[\psi(\omega, \theta_0) \psi^{\intercal}(\omega, \theta_0)]$ and $A_0 = E[H(\omega, \theta_0)].$ Let $g(\beta, \gamma) = \sum_{k=1}^K p(k\mid x)v^\intercal_k(x)\beta_k$, we have
$$
\sqrt{n}(\hat{\tau}(x) - \tau(x)) = \sqrt{n}\left(g(\hat{\beta}, \hat{\gamma})-g(\beta, \gamma)\right) \overset{d}{\rightarrow} N(0, (\mathcal{D}g)A^{-1}_0B_0A^{-1}_0(\mathcal{D}g)^\intercal)
$$
by the multivariate delta method, where $\mathcal{D}g = \begin{bmatrix}
   \frac{\partial^\intercal g(\beta, \gamma)}{\partial \beta} & \frac{\partial^\intercal g(\beta, \gamma)}{\partial \gamma}
\end{bmatrix}^\intercal$ with derivatives $\frac{\partial g(\beta, \gamma)}{\partial \beta_k} = p(k\mid x)v^\intercal_k(x)$ for $k = 1, \ldots, K$ and $\frac{\partial g(\beta, \gamma)}{\partial \gamma_k} = v^\intercal_k(x)\beta_k x(\zeta_k - p(k\mid x))$ for $k = 1, \ldots, K - 1$. 
\\
\hspace{14.3cm} $\square$

\end{proof}

\noindent \begin{proof}(Theorem 2) Suppose the membership probabilities $p(1\mid x) = p(2\mid x) = p$ for all $ x$, and $E[\epsilon^2\mid X, A] = \sigma^2.$ By Lemma 1,   Theorem 4.2 (Pointwise Normality) in \cite{belloni2015some} and Assumptions 8 and 9, we have that for any $\alpha,$
$$\sqrt{n} \alpha^{\intercal}(\hat{\beta}-\beta)\overset{d}{\rightarrow} N\left(0,\norm{s(x)}^2_2\right),$$
where $s(x) = \Omega^{1/2}\alpha$, $\Omega = \Gamma^{-1} E[\epsilon_i^2 u_iu_i^\intercal]\Gamma^{-1},$ and $\Gamma = E(u_iu_i^{\intercal})$. Let $\alpha = Z(x)v(x)$, then it follows that 
$$\sqrt{n}(\hat{\tau}(x)-\tau(x)) \overset{d}{\rightarrow} N\left(0, \norm{s(x)}^2_2\right).$$
The asymptotic variance of the multi-study $R$-learner estimator $\hat{\tau}(x)$ is
\begin{align*}
       var(\hat{\tau}(x))_{\text{Asy}} &= \frac{1}{n}v^\intercal (x) Z(x) \Omega Z(x)v(x)\\
       &= \frac{1}{n} v^\intercal (x) Z(x) \Gamma^{-1}E[\epsilon^2uu^\intercal]\Gamma^{-1}Z(x)v(x)\\
       &= \frac{\sigma^2}{n}  v^\intercal (x) Z(x) \Gamma^{-1}Z(x)v(x)\\
       &= \frac{\sigma^2p^2}{n}v^\intercal(x)\Gamma^{-1}v(x) \label{S1}\tag{S1}
\end{align*}

\noindent To obtain the asymptotic variance of the study-specific $R$-learner estimator, i.e., $$\hat{\tau}^{SS}(x) = \sum_{k=1}^K p v_k^\intercal(x) \hat{\beta}^R_k,$$ where $\hat{\beta}^R_k$ is obtained by fitting an $R$-learner to study $k$,
$$ \hat{\beta}^R_k = \argmin_b \frac{1}{n_k}\sum_{i=1}^{n_k} \left[\{Y_{i}-\hat{m}_k^{-q_k(i)}(X_{i})\}-(u^R_{k,i})^{\intercal}b\right]^2 + O_p(a_n^2),$$
and $u^R_{k,i} = p^{-1}u_{k,i}$, we
use the same strategy as above. Let $\Omega^R_k = (\Gamma^R_{kk})^{-1}E[\epsilon_{k,i}^2 u^R_{k,i}(u^R_{k,i})^\intercal](\Gamma^R_{kk})^{-1}$. The asymptotic variance of the study-specific $R$-learner estimator is  
\begin{align*}
    var(\hat{\tau}^{SS}(x))_{\text{Asy}} 
    &= \sum_{k=1}^K \frac{1}{n_k} p^2v^\intercal_k(x)\Omega_k^Rv_k(x)\\
    &= \sum_{k=1}^K \frac{1}{n_k}p^2v^\intercal_k(x)(\Gamma^R_{kk})^{-1}E[\epsilon_k^2 u^R_{k}(u^R_{k})^\intercal](\Gamma^R_{kk})^{-1}v_k(x)\\
     &= \sum_{k=1}^K \frac{\sigma^2_k}{n_k}p^2v^\intercal_k(x)(\Gamma^R_{kk})^{-1}v_k(x)\\
      &= \sum_{k=1}^K \frac{\sigma^2_k}{n_k}p^2v^\intercal_k(x)\left(E\left[\frac{1}{p^2}u_ku_k^\intercal\right]\right)^{-1}v_k(x)\\
       &= \sum_{k=1}^K \frac{\sigma^2_k}{n_k}p^4v^\intercal_k(x)\left(E\left[u_ku_k^\intercal\right]\right)^{-1}v_k(x) \\
       &= \frac{p^2}{n}v^\intercal(x)\text{blkdiag}\left(\frac{n\sigma^2_kp^2}{n_k}\Gamma^{-1}_{kk}\right)v(x) \label{S2}\tag{S2}
\end{align*}

\noindent The goal is to compare (\ref{S1}) with (\ref{S2}). We want to show that for all $x \in \mathcal{X},$

\begin{align*}
    &\frac{p^2}{n}v^\intercal(x)\left(\text{blkdiag}\left(\frac{n\sigma^2_kp^2}{n_k}\Gamma^{-1}_{kk}\right) - \sigma^2\Gamma^{-1}\right)v(x) > 0 \tag{S3} \label{S3}
    \end{align*}

\noindent For $K = 2,$ we have  $$\Gamma = E(u_iu_i^\intercal) = E\begin{bmatrix} u_{1,i}u_{1, i}^\intercal & u_{1,i}u_{2,i}^\intercal \\
u_{2,i}u_{1,i}^\intercal & u_{2,i}u_{2,i}^\intercal
\end{bmatrix} = \begin{bmatrix} \Gamma_{11} & \Gamma_{12} \\
\Gamma_{21} & \Gamma_{22} 
\end{bmatrix}.$$  
Let $v_k(\cdot) = v^*(\cdot)$ for all $k,$ then 
\begin{align*}
    \Gamma_{11} &= E[(A_i - e_1(X_i))^2p^2v^*(X_i)(v^*(X_i))^\intercal]\\
    &= E[e_1(X_i)(1-e_1(X_i))p^2v^*(X_i)(v^*(X_i))^\intercal]\\
        \Gamma_{22} &= E[(A_i - e_2(X_i))^2p^2v^*(X_i)(v^*(X_i))^\intercal]\\
    &= E[e_2(X_i)(1-e_2(X_i))p^2v^*(X_i)(v^*(X_i))^\intercal]\\
    \Gamma_{12} = \Gamma_{21} &= E[(A_i - e_1(X_i))(A_i-e_2(X_i))p^2v^*(X_i)(v^*(X_i))^\intercal]\\
    &= p(\Gamma_{11} + \Gamma_{22})
\end{align*}
Suppose $e_1(\cdot)(1 - e_1(\cdot)) = ce_2(\cdot)(1-e_2(\cdot))$ for some positive constant $c,$ then

$$\Gamma = \begin{bmatrix} \Gamma_{11} & p\Gamma_{11}(1 + c) \\
p\Gamma_{11}(1 + c) & c\Gamma_{11} 
\end{bmatrix}.$$
Its inverse is 
$$\Gamma^{-1} = \Sigma = \begin{bmatrix} \Sigma_{11} & \Sigma_{12} \\
\Sigma_{21} & \Sigma_{22} 
\end{bmatrix}$$
where \begin{align*}
    \Sigma_{11} &= (\Gamma_{11}-p^2(1+c)^2 \Gamma_{11}(c\Gamma_{11})^{-1}\Gamma_{11})^{-1}\\
    &= \frac{c}{c-p^2(1+c)^2}\Gamma_{11}^{-1}\\
    \Sigma_{22} &= (c\Gamma_{11} - p^2(1+c)^2\Gamma_{11}\Gamma^{-1}_{11}\Gamma_{11})^{-1}\\
    &= \frac{1}{c-p^2(1+c)^{2}}\Gamma^{-1}_{11}\\
    \Sigma_{12} = \Sigma_{21} &= \frac{-c}{c-p^2(1+c)^2}\Gamma_{11}^{-1}(p(1 + c)\Gamma_{11}(c\Gamma_{11})^{-1})\\
    &= \frac{-p(1+c)}{c-p^2(1+c)^2}\Gamma^{-1}_{11}
\end{align*}
To establish (\ref{S3}), we want to show that 
$$\text{blkdiag}\left(\frac{n\sigma^2_kp^2}{n_k}\Gamma^{-1}_{kk}\right)-\sigma^2\Gamma^{-1} = \begin{bmatrix}
        \left(\frac{n\sigma^2_1p^2}{n_1}-\frac{\sigma^2c}{c-p^2(1+c)^2}\right) \Gamma_{11}^{-1} & \frac{p(1+c)}{c-p^2(1+c)^2}\Gamma^{-1}_{11}\\
        \frac{p(1+c)}{c-p^2(1+c)^2}\Gamma^{-1}_{11} & \left(\frac{n\sigma^2_2p^2}{n_2c}-\frac{\sigma^2}{c-p^2(1+c)^2}\right)\Gamma_{11}^{-1}
    \end{bmatrix}$$ is positive definite. Since $\Gamma$ is positive definite, $\Gamma^{-1}_{11}$ is also positive definite. For $c \neq 1,$ $\frac{n\sigma^2_2p^2}{n_2c}-\frac{\sigma^2}{c-p^2(1+c)^2}$ is a positive constant, so $\left(\frac{n\sigma^2_2p^2}{n_2c}-\frac{\sigma^2}{c-p^2(1+c)^2}\right)\Gamma^{-1}_{11}$ is positive definite. It remains to show that the Schur complement, $$\left(\frac{n\sigma^2_1p^2}{n_1}-\frac{\sigma^2c}{c-p^2(1+c)^2}\right) \Gamma_{11}^{-1} - \left(\frac{p(1+c)}{c-p^2(1+c)^2}\right)^2\left(\frac{n_2c(c-p^2(1+c)^2)}{n\sigma^2_2p^2(c-p^2(1+c)^2) - n_2c\sigma^2}\right)\Gamma^{-1}_{11}$$ is positive definite. This is equivalent to showing 
    
    \begin{small}
    \begin{align*}
    &\frac{n\sigma^2_1p^2(c - p^2(1 + c)^2) - \sigma^2cn_1}{n_1(c - p^2 (1 + c)^2)}-\frac{p^2(1+c)^2}{c-p^2(1+c)^2}\left(\frac{n_2c}{n\sigma^2_2p^2(c-p^2(1+c)^2) - n_2c\sigma^2}\right) > 0\\
    \Longleftrightarrow \qquad & \frac{n\sigma^2_1p^2(c - p^2(1 + c)^2) - \sigma^2cn_1}{n_1} < \frac{n_2cp^2(1+c)^2}{n\sigma^2_2p^2(c-p^2(1+c)^2) - cn_2\sigma^2} \quad (\text{Because } c < p^2(1+c)^2)\\
      \Longleftrightarrow \qquad & (n\sigma^2_1p^2(c - p^2(1 + c)^2) - \sigma^2cn_1)(n\sigma^2_2p^2(c - p^2(1 + c)^2) - \sigma^2cn_2) > n_1n_2cp^2(1+c)^2\\
       \Longleftrightarrow \qquad & n^2\sigma^2_1\sigma_2^2p^4(c - p^2(1+c)^2)^2 - (n_1\sigma^2_2 + n_2\sigma^2_1)n\sigma^2p^2c(c-p^2(1 + c)^2) > n_1n_2cp^2(1+c)^2 - \sigma^4c^2n_1n_2\\
        \Longleftrightarrow \qquad & \sigma^2_1\sigma^2_2(np^2(c - p^2(1+c)^2))^2 - (n_1\sigma^2_2 + n_2 \sigma^2_1)n\sigma^2p^2c(c-p^2(1 + c)^2) > -n_1n_2c(\sigma^4c - p^2(1+c)^2)
\end{align*}
    \end{small}
\noindent Note that the first term is positive and the second is negative. So the left hand side of the inequality is positive. Assuming $\sigma^2 > (1 + c)(2 \sqrt{c})^{-1}$ for $c \neq 1,$ the right hand side is negative. $\square$ 

\end{proof}

\label{lastpage}

\end{document}